\documentclass[american,english]{IEEEtran}
\usepackage[T1]{fontenc}
\usepackage[latin9]{inputenc}
\usepackage{amsmath}
\usepackage{amssymb}
\usepackage{graphicx}

\makeatletter

\providecommand{\tabularnewline}{\\}

\usepackage{cite}

\renewcommand{\citedash}{\citeright\hbox{--}\penalty\@m\citeleft}
\renewcommand{\citepunct}{\citeright,\penalty\@m\hskip.13emplus.1emminus.1em\citeleft}

\makeatother

\usepackage{babel}

\usepackage{url}

\makeatother

\usepackage{babel}
\begin{document}

\title{Spectral Efficiency Optimization in Flexi-Grid Long-Haul Optical
Systems}

\author{Tommaso Foggi, Giulio Colavolpe \IEEEmembership{Senior Member, IEEE},
Alberto Bononi \IEEEmembership{Senior Member, IEEE}, and Paolo Serena\foreignlanguage{american}{
\IEEEmembership{Member, IEEE}\thanks{T. Foggi is with CNIT Research Unit, I-43124 Parma, Italy. G. Colavolpe, A. Bononi and P. Serena are with Dip. Ing. Informazione, Università degli Studi di Parma, I-43124 Parma, Italy.}
\thanks{The paper was submitted in part at the IEEE International Conference on Communications (ICC 2015), London, UK, June 2015}.}}
\maketitle
\begin{abstract}
Flexible grid optical networks allow a better exploitation of fiber
capacity, by enabling a denser frequency allocation. A tighter channel
spacing, however, requires narrower filters, which increase linear
intersymbol interference (ISI), and may dramatically reduce system
reach. Commercial coherent receivers are based on symbol by symbol
detectors, which are quite sensitive to ISI. In this context, Nyquist
spacing is considered as the ultimate limit to wavelength-division
multiplexing (WDM) packing. 

In this paper, we show that by introducing a limited-complexity trellis
processing at the receiver, either the reach of Nyquist WDM flexi-grid
networks can be significantly extended, or a denser-than-Nyquist channel
packing (i.e., a higher spectral efficiency (SE)) is possible at equal
reach. By adopting well-known information-theoretic techniques, we
design a limited-complexity trellis processing and quantify its SE
gain in flexi-grid architectures where wavelength selective switches
over a frequency grid of 12.5GHz are employed.\end{abstract}


\section{Introduction}

Spurred by the relentless increase of data traffic, coherent optical
systems were revived in the last decade, and many different paths
were undertaken in order to better exploit the huge capacity of the
fiber channel, from polarization-multiplexing (PM) and high-order
modulations, to dense wavelength division multiplexing (WDM) solutions
like Nyquist-WDM or orthogonal frequency division multipleximg (OFDM)
\cite{BoCuCaPoFo11,jansen:OFT09}. Although to cope with the foreseen
capacity crunch of the existing fiber infrastructure \cite{Chraplyvy10}
the long-term solution will likely be the use of multimode fibers
and multi-input-multi-output processing -- a solution that requires
replacing the existing fiber infrastructure -- yet a less disruptive
step towards increasing SE has been the introduction of flexi-grid
WDM networks \cite{ITU12} where throughput increase is achieved through
the reduction of channel spacing. However, the deployment of new generation
flexible wavelength selective switches (WSS) \cite{Finisar12} in
reconfigurable optical add-and-drop multiplexers (ROADM), compatible
with the aforemoentioned standard, entails a careful redesign of optical
systems, and specifically transmission and reception techniques. In
fact, the presence of WSS with 12.5 GHz granularity prevents the SE
increase through the simple adoption of Nyquist-shaped signaling,
as the effect of crossing ROADMs, and therefore cascaded WSS filtering,
is detrimental even after a few nodes \cite{MoreaOFC14}. 

We employ a maximum-a-posteriori (MAP) symbol detector \cite{BaCoJeRa74}
with a minimum number of states in order to mitigate the effects of
WSS inline filtering, and compare it to the conventional symbol-by-symbol
detector. The comparison is carried out for a polarization-multiplexed
quaternary phase shift keying (QPSK) modulation, both in terms of
pre forward-error-correction (FEC) bit error rate (or equivalently
Q-factor) and in terms of achievable information rate (AIR), and thus
achievable SE, by resorting to the simulation-based technique detailed
in \cite{CoFo14}. While AIR represents a theoretical value that may
be achieved by some optimal FEC, we next extend the analysis by equipping
our receivers with existing low-density parity check codes (LDPC)
designed for satellite links \cite{DVB-S2-TR}, and check their SE
against that obtained from AIR. The SE of such practical LDPCs turns
out to be close to the value of the achievable SE, thus providing
evidence of the practical meaning of the achievable SE as an upper
bound on SE of all FECs working with such a receiver. We prove, for
instance, that even 2-state MAP symbol detectors%
\footnote{Notice that our receiver processes the PM-QPSK signal components separately,
thus it entails four 2-state MAP detectors.%
} allow to more than double the maximum reach when using 25\% overhead
FECs and a Nyquist channel spacing with tight optical filtering. Next,
with the same MAP receiver, we show that it is possible to further
increase the system SE by going beyond the Nyquist limit \cite{CoFoMoPi11},
hence transmitting at a higher symbol rate at the same spacing. Results
show that, taking the standard symbol-by-symbol threshold detector
at the lowest symbol rate as a reference, a gain up to almost 50\%
in SE (i.e., 50\% more throughput) is possible. 

The paper is organized as follows. In Section \ref{sec:System-model},
a description of the system model link architecture and the adopted
detection strategy are provided. Section \ref{sec:Spectral-Efficiency-Analysis}
gives insight on the theoretical approach chosen to evaluate the receiver
performance and to provide an analysis of system impairments. Section
\ref{sec:Numerical-Results} reports on the numerical results and
the observations that arise. Finally, conclusions are drawn in Section
\ref{sec:Conclusions}.

\section{System model\label{sec:System-model}}

The considered system is closely related to that described in \cite{MoreaOFC14}.
In our WDM simulated system, $N_{c}$ PM linearly-modulated signals
are launched on random polarizations and with random detuning with
respect to central frequencies. In the following, we will consider
a QPSK modulation format on each carrier and each polarization. The
general expression for the complex envelope of the signal transmitted
on the $\ell$th carrier and the $i$th polarization ($i=1,2$) is
\begin{equation}
\sum_{k=0}^{K-1}x_{k}^{(i,\ell)}p(t-kT-\tau^{(i,\ell)})e^{j[2\pi\ell(F+\Delta_{\ell})t+\theta^{(i,\ell)}]}\label{eq:tx_signal}
\end{equation}
where $p(t)$ is the shaping pulse having root raised-cosine (RRC)
spectrum with roll-off $\alpha=0.1$ (obtained by proper transmit-side
electrical filtering), $K$ the number of symbols transmitted over
each carrier and each polarization, $T$ the symbol interval, $x_{k}^{(i,\ell)}$
the symbol transmitted over the $\ell$th carrier of the $i$th polarization
during the $k$th symbol interval, $\tau^{(i,\ell)}$ and $\theta^{(i,\ell)}$
the delay and the initial phase of the $i$th polarization and $\ell$th
carrier, respectively, $F$ the frequency spacing between two adjacent
carriers, and $\Delta_{\ell}$ the possible frequency offset (small
compared to the frequency spacing).%
\footnote{The frequency offset can be considered to be two order of magnitude
smaller than the carrier spacing. In our simulations we included random
offsets in the range $\pm1$\% of the carrier spacing.%
} The transmitted symbols were obtained from a stream of information
bits, by properly encoding with a binary FEC and Gray mapped onto
the QPSK constellation.

The transmitted signal was then launched into a dispersion unmanaged
fiber link with variable number of spans $N_{s}$, characterized by
the presence of ROADM nodes, one every two fiber spans. Therefore,
the number of crossed ROADMs was equal to half the number of spans.
Each span had 120 km of single mode fiber (SMF) and an erbium-doped
fiber amplifier (EDFA) with a noise figure of 6 dB. Since ROADMs are
here intended as simple pass-through nodes, they basically just introduce
the filtering effect of two WSS, modeled as 3rd-order super-Gaussian
filters. The bandwidth of such filters can be determined once the
flexible grid spacing has been selected. We fixed the channel spacing
to 37.5 GHz, which implies a 3-dB filter bandwidth of 35.75 GHz \cite{Finisar12,MoreaOFC14}. 

\begin{figure}
\begin{centering}
\includegraphics[width=0.95\columnwidth]{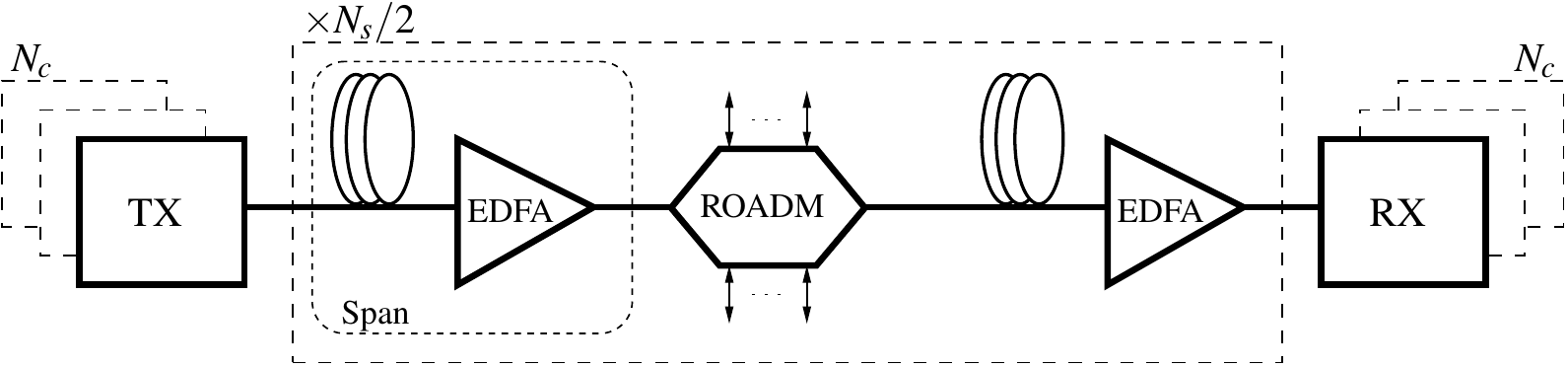}
\par\end{centering}

\protect\caption{\label{fig:system_model}Schematic of the simulated optical link.}

\end{figure}

Fiber propagation was impaired by group velocity dispersion (GVD),%
\footnote{We also considered polarization mode dispersion (PMD) with values
of the differential group delay of typical fibers and noted no performance
difference. Thus, PMD is not present in current results. %
} and nonlinear effects. These latter effects were simulated by the
split step Fourier method (SSFM) \cite{Men03} applied to the Manakov
nonlinear equation with proper step size.%
\footnote{We optimized the step-size for each launch power, by increasing the
value in trial simulations until we noted no performance variation.%
} The symbol rate $R$ of each signal was initially fixed to 32.5 Gbaud,
as in \cite{MoreaOFC14}. Then, in order to demonstrate the advantages
of the time-frequency packing (TFP) technique \cite{CoFo14}, we increased
the data rate beyond the Nyquist limit, up to 75 Gbaud, while keeping
all remaining parameters unchanged (i.e., we did not change the filter
bandwidths). Fig.~\ref{fig:system_model} shows the block diagram
of the generic simulated link.

At the receive side, coherent detection was performed \cite{CoFoFoPr08}.
The received optical field was first filtered by a 4th-order super-Gaussian
filter having a 3-dB bandwidth of 35.75 GHz, which allows to select
the desired channel, and then converted to the electrical domain through
a $90^{\mathrm{o}}$ optical hybrid. Digital signal processing (DSP)
was then performed, as explained in detail in \cite{CoFo14}. After
sampling, compensation of the cumulated GVD was performed by two fixed-tap
equalizers (one per polarization) and then frame and frequency synchronization
and compensation were performed. A two-dimensional fractionally-spaced
adaptive minimum mean square error (MMSE) feed-forward equalizer (FFE)
performed compensation of the residual GVD and polarization mode dispersion
(PMD), and also performed polarization demultiplexing. The number
of taps was chosen sufficiently high so that GVD and PMD did not entail
any penalty, whereas a coefficient adaptation step-size value of $10^{-3}$
came up to be optimal in any case. Finally phase noise was tracked
by a proper decision-directed phase estimation and compensation module.
All synchronization aspects were neglected here---perfect synchronization
was assumed. These aspects will be discussed in detail in a future
paper. Finally, signal samples fed the detector, which, in the MAP
case, iteratively exchanges soft information with the LDPC decoder.

We considered two kinds of detectors. First, we used a conventional
symbol-by-symbol detector/demapper that neglects channel memory (i.e.,
the detector commonly used in coherent receivers). As a second more
sophisticated solution, we employed a MAP symbol detector with a minimum
number of states (see Section \ref{sec:Numerical-Results}), preceeded
by a channel shortener \cite{RuPr12}, which is essentially a linear
filter with a few taps, and whose computation is based on the estimation
of the overall channel impulse response. The shortener helps coping
with the intersymbol interference (ISI) not accounted for by the limited
detector memory, as explained in \cite{CoFo14}. More details on the
simulated system and receiver parameters will be given in Section
\ref{sec:Numerical-Results}. 

On the described system, we performed two kinds of simulations. First,
we evaluated the achievable SE, as explained in the next section,
by varying the launch power per channel and number of spans $N_{s}$
of the link. Then, we also employed a set of codes with rates in the
range 1/3-9/10. These codes were used to confirm the AIR semi-analytical
predictions.

\section{Spectral Efficiency Analysis\label{sec:Spectral-Efficiency-Analysis}}

We now describe the framework used to evaluate the performance limits
of all the optical transmission systems considered in this paper. 

We consider an optical channel with linear and nonlinear distortions,
simulated through the SSFM. Denoting by $\mathbf{y}$ a proper discrete-time
sufficient statistic used for detection of the information symbols
$\mathbf{x}=\{x_{k}^{(i,\ell)}\}_{k,i,\ell}$, the information rate
(IR), i.e., the average mutual information when the information symbols
are independent and uniformly distributed random variables belonging
to the given constellation, is defined as%
\footnote{The factor 2 takes into account the presence of two polarizations.%
} 
\begin{equation}
I(\mathbf{x};\mathbf{y})\!=\!\!\!\lim_{K\rightarrow\infty}\frac{1}{2N_{c}K}E\Bigg\{\log_{2}\frac{p(\mathbf{y}\arrowvert\mathbf{x})}{\sum_{\mathbf{x'}}p(\mathbf{y}\arrowvert\mathbf{x'})P(\mathbf{x'})}\Bigg\}\negthickspace\left[\frac{\mbox{bit}}{\mbox{ch. use}}\right]\label{e:IR-1}
\end{equation}
where $E[\cdot]$ denotes expectation, $p(\cdot)$ a probability density
function (PDF) and $P(\cdot)$ a probability mass function (PMF).
The SE is the IR per unit bandwidth and unit time
\[
\mathrm{SE=}\frac{I(\mathbf{x};\mathbf{y})}{FT}\quad[\mbox{b/s/Hz}]
\]
since $FT$ is the time-frequency slot devoted to the transmission
of symbol $x_{k}^{(i,\ell)}$.

The computation of IR and SE requires the availability of the pdfs
$p(\mathbf{y}\arrowvert\mathbf{x})$ and $p(\mathbf{y})=\sum_{\mathbf{x'}}p(\mathbf{y}\arrowvert\mathbf{x'})P(\mathbf{x'})$.
However, they are not known in closed form, nor can we resort to the
simulation method in \cite{ArLoVoKaZe06} to compute them. In fact,
this method requires that the channel at hand has a finite memory
and the availability of an optimal detector for it \cite{ArLoVoKaZe06}.
These conditions are clearly not satisfied in our scenario \cite{EssTka12,Essjlt10}.
We may thus resort to the computation of a proper lower bound of the
IR (and thus of the SE) obtained by substituting $p(\mathbf{y}\arrowvert\mathbf{x})$
in (\ref{e:IR-1}) with an arbitrary auxiliary channel law $q(\mathbf{y}\arrowvert\mathbf{x})$
with the same input and output alphabets as the original channel (mismatched
detection~\cite{MeKaLaSh94,ArLoVoKaZe06,CoFoMoPi11,Serena12}). The
resulting lower bound reads as
\begin{equation}
I_{\mathrm{LB}}(\mathbf{x};\mathbf{y})=\lim_{K\rightarrow\infty}\frac{1}{2N_{c}K}E\Bigg\{\log_{2}\frac{q(\mathbf{y}\arrowvert\mathbf{x})}{\sum_{\mathbf{x'}}q(\mathbf{y}\arrowvert\mathbf{x'})P(\mathbf{x'})}\Bigg\}.\label{e:LB}
\end{equation}
If the auxiliary channel law is representative of a finite-state channel,
pdfs $q(\mathbf{y}\arrowvert\mathbf{x})$ and $q_{p}(\mathbf{y})=\sum_{\mathbf{x'}}q(\mathbf{y}\arrowvert\mathbf{x'})P(\mathbf{x'})$
can be computed, this time, by using the optimal MAP symbol detector
for that auxiliary channel \cite{ArLoVoKaZe06}. Such a detector,
which will clearly be suboptimal for the actual channel, will have
at its input the sequence $\mathbf{y}$ generated by simulation \emph{according
to the actual channel model,} and the expectation in\emph{ }(\ref{e:LB})
is meant with respect to the input and output sequences generated
accordingly \cite{ArLoVoKaZe06}. Thus, no assumption on the true
statistics of the discrete-time received sequence is required for
the design of the adopted detector, since it is designed for the auxiliary
channel. Similarly, the true statistics of the sequence $\mathbf{y}$
are not analytically required for its generation, since they can be
obtained by SSFM simulation through the actual nonlinear channel.
If we change the auxiliary channel (and thus the trellis metrics based
on it) we obtain different lower bounds on the information rate but,
in any case, such bounds are \emph{achievable} by those receivers,
according to mismatched detection~\cite{MeKaLaSh94,ArLoVoKaZe06}.
We will thus say, with an abuse of terminology, that the computed
lower bounds are the maximum SE values of the considered channel when
those receivers are employed. All these considerations hold for any
actual channel, including nonlinear and non-Gaussian ones.

This technique thus allows to evaluate the AIR for receivers of reduced
complexity. In fact, it is sufficient to consider an auxiliary channel
which is a simplified version of the actual channel in the sense that
only a portion of the true channel memory and/or a limited number
of impairments are present. The considered receivers have been described
in the previous section. As mentioned, we have assumed that parallel
independent detectors are employed, one for each carrier and each
polarization. In other words, intercarrier interference (ICI) is not
coped with at the receiver, since multiuser detection is considered
too computationally demanding. This corresponds to the adoption of
an auxiliary channel model that can be factorized into the product
\[
q(\mathbf{y}\arrowvert\mathbf{x})=\prod_{i}\prod_{\ell}q(\mathbf{y}^{(i,\ell)}\arrowvert\mathbf{x}^{(i,\ell)})
\]
where $\mathbf{y}^{(i,\ell)}$ is a proper discrete-time received
sequence used for detection of symbols $\mathbf{x}^{(i,\ell)}=\{x_{k}^{(i,\ell)}\}$
transmitted over the $\ell$th carrier and the $i$th polarization.
Under this assumption, we simply have 
\begin{equation}
I_{\mathrm{LB}}(\mathbf{x}^{(i,\ell)};\mathbf{y}^{(i,\ell)})=\lim_{K\rightarrow\infty}\frac{1}{K}E\Bigg\{\log_{2}\frac{q(\mathbf{y}^{(i,\ell)}\arrowvert\mathbf{x}^{(i,\ell)})}{q_{p}(\mathbf{y}^{(i,\ell)})}\Bigg\}\,,\label{e:IR-2}
\end{equation}
i.e., the result can be computed by considering only one carrier.
In a practical scenario with a finite number of carriers, we will
consider the central carrier only, thus avoiding the computation on
the border carriers which are affected by a smaller amount of ICI,
thus obtaining a further lower bound. 

Note that, as stated, we are not able to compute the IR of the true
channel since the optimal receiver is unknown and possibly of unmanageable
complexity. We take here the pragmatic approach of considering only
limited-complexity suboptimal receivers. For such receivers we are
indeed able to compute the relevant IR which will be called \textit{achievable}
IR. The corresponding achievable lower bound on SE (\textit{achievable}
SE in the following) is thus 
\begin{equation}
\eta_{\mathrm{LB}}=\frac{1}{FT}I_{\mathrm{LB}}(\mathbf{x}^{(i,\ell)};\mathbf{y}^{(i,\ell)})\quad[\mbox{\textrm{b/s}/\textrm{Hz}}].\label{e:eta}
\end{equation}

The auxiliary channel that we adopted for the MAP symbol detector
design neglects the presence of channel nonlinear effects, and assumes
that GVD and PMD have been perfectly compensated. Basically, the detector
is designed by taking into account transmit side shaping pulse and
inline filtering, so that the sufficient statistics $\mathbf{y^{\mathrm{(\mathit{i},\ell)}}}$
can be obtained by sampling the output of a filter matched to the
received pulse in the absence of GVD, PMD, and nonlinear effects,
i.e., the transmit pulse after cascaded inline filtering.%
\footnote{The FFE taps are designed by using the matched filter output as the
target response, so that the equalizer does not remove the ISI induced
by filtering but only performs matched filtering. It is worth noting
that, if extremely narrow optical filtering is employed at the receive
side, the electrical compensation of chromatic dispersion through
the non-adaptive equalizers may be inaccurate. In this case, a wider
optical filter can be used, compatibly with the system design, in
order to leave the useful component of the received signal unchanged,
whereas matched filtering is implemented by the adaptive equalizer. %
} In SSFM simulations, noise contributions introduced by EDFAs are
added along the whole optical link at each span. Hence inline filtering
has an incremental effect on the propagating signal, whereas for the
auxiliary channel we assume that all noise is added at the end of
the link. Thus, at the receive side it is possible to estimate the
overall channel response (e.g., through a simple MMSE estimator) without
any knowledge on the link configuration, which corresponds to the
most practical way to design the MAP symbol detector. Notice that
the presence of other conventional receive-side filters, with bandwidth
compatible with the chosen frequency grid, does not imply changes
to the established matched filter response, thus does not affect the
aforementioned considerations. Given this auxiliary channel law, the
optimal MAP symbol detector is described in \cite{CoBa05b} (see \cite{CoFo14}
for more details). 

As a concluding remark, we would like to point out that this technique
allows to compute the achievable limit of the considered receivers
without taking into account specific coding schemes, being understood
that, with a properly designed channel code, the information-theoretic
performance can be closely approached. Section \ref{sec:Numerical-Results}
will report some design cases for these codes with the aim of showing
that, indeed, the performance predicted by the achievable SE can be
approached. 

\begin{table}
\begin{centering}
{\scriptsize{}}%
\begin{tabular}{|c|c|c|c|c|}
\hline 
 & {\scriptsize{}TX Optical} & {\scriptsize{}Inline WSS} & {\scriptsize{}RX Optical} & {\scriptsize{}RX electrical}\tabularnewline
\hline 
{\scriptsize{}Type} & {\scriptsize{}RRC, $\alpha$=0.1} & {\scriptsize{}3rd Gauss.} & {\scriptsize{}4th Gauss.} & {\scriptsize{}5th Bessel}\tabularnewline
\hline 
{\scriptsize{}Bandwidth {[}GHz{]}} & {\scriptsize{}35.75} & {\scriptsize{}35.75} & {\scriptsize{}35.75} & {\scriptsize{}16}\tabularnewline
\hline 
\end{tabular}
\par\end{centering}{\scriptsize \par}

\vspace{2mm}

\protect\caption{\label{tab:filt}Filters parameters.\textbf{ }}
\end{table}

\section{Numerical Results\label{sec:Numerical-Results}}

In this section, we present the simulation results of the optical
channel described in Section \ref{sec:System-model}. The WDM input
signal in (\ref{eq:tx_signal}) had 11 channels with spacing 37.5
GHz and was launched in the line with $P$ power per channel and propagated
through a dispersion-uncompensated (DU) link of $N_{s}$ identical
120 km single-mode fiber (SMF) spans. Fiber parameters include dispersion
16.63 $\mathrm{ps/nm/km}$, attenuation 0.23 $\mathrm{dB/km}$, and
nonlinear index $\gamma=$1.3 $\mathrm{W^{-1}km^{-1}}$. Every two
spans we included a ROADM, whose only effect is a filtering due to
the presence of two WSS on the signal path. In Tab.~\ref{tab:filt},
we recall system filter types and bandwidths, which are kept unchanged
in all simulations. The effect of cascading WSS is then summarized
in Tab.~\ref{tab:Power_loss}, which reports the power loss of the
RRC-filtered 32.5 Gbaud signal versus the number of crossed WSS.

\begin{figure}
\begin{centering}
\includegraphics[angle=-0,width=0.95\columnwidth]{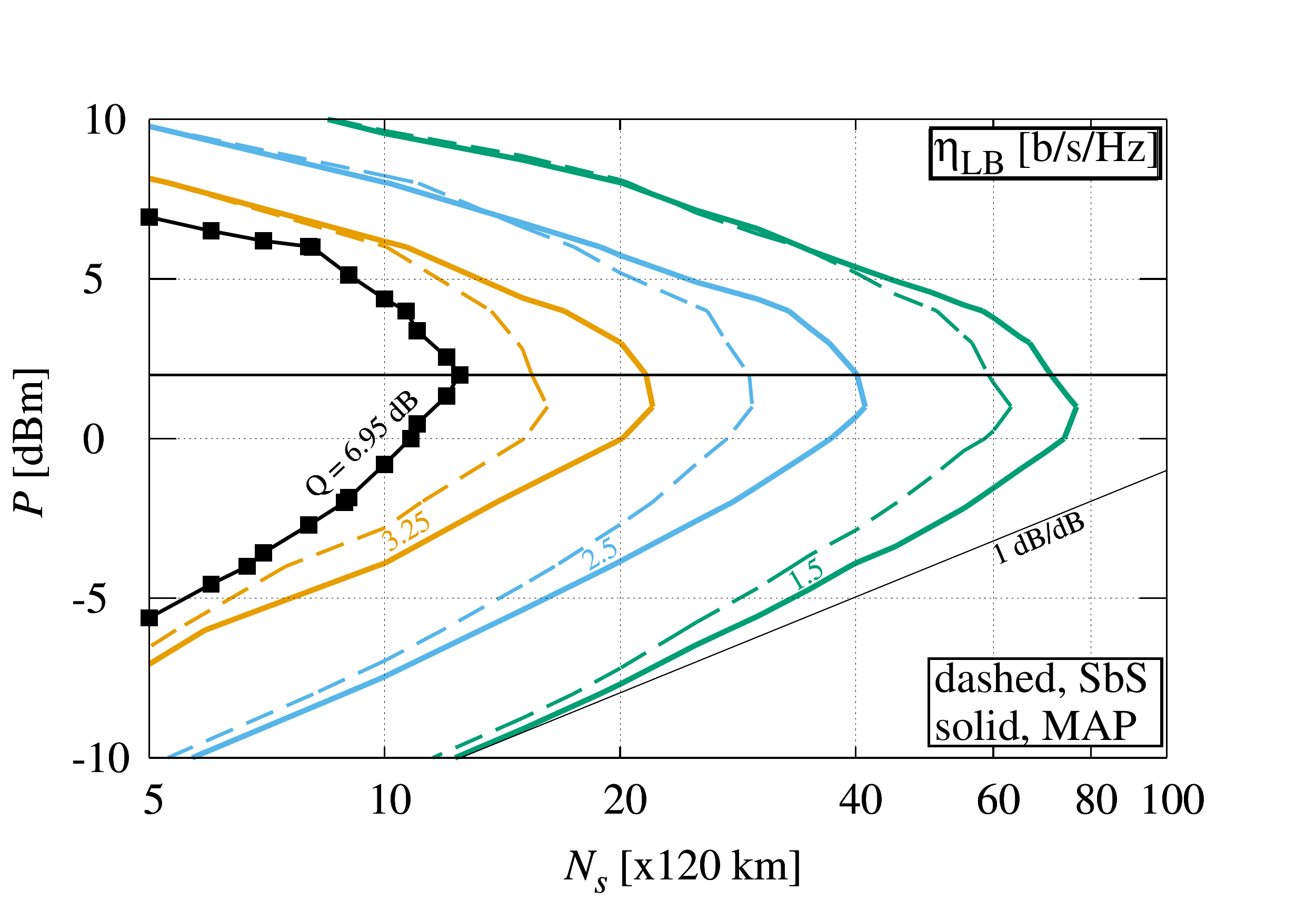}
\par\end{centering}

\protect\caption{\label{fig:grid32.5}Horizontal cuts of spectral efficiency versus
launch power per channel and number of spans $N_{s}$, in a $N_{s}\times120$
km SMF DU link, with $R$=32.5 Gbaud, $F$=37.5 GHz, MAP detector
with $L$=1 and conventional symbol-by-symbol detector. It is also
reported the horizontal cut corresponding to BER=0.0132, or Q-factor=6.95
dB, assumed as a pre-FEC threshold for 25\% overhead code, symbol-by-symbol
detector.}
\end{figure}

\begin{figure}
\begin{centering}
\includegraphics[angle=-0,width=0.95\columnwidth]{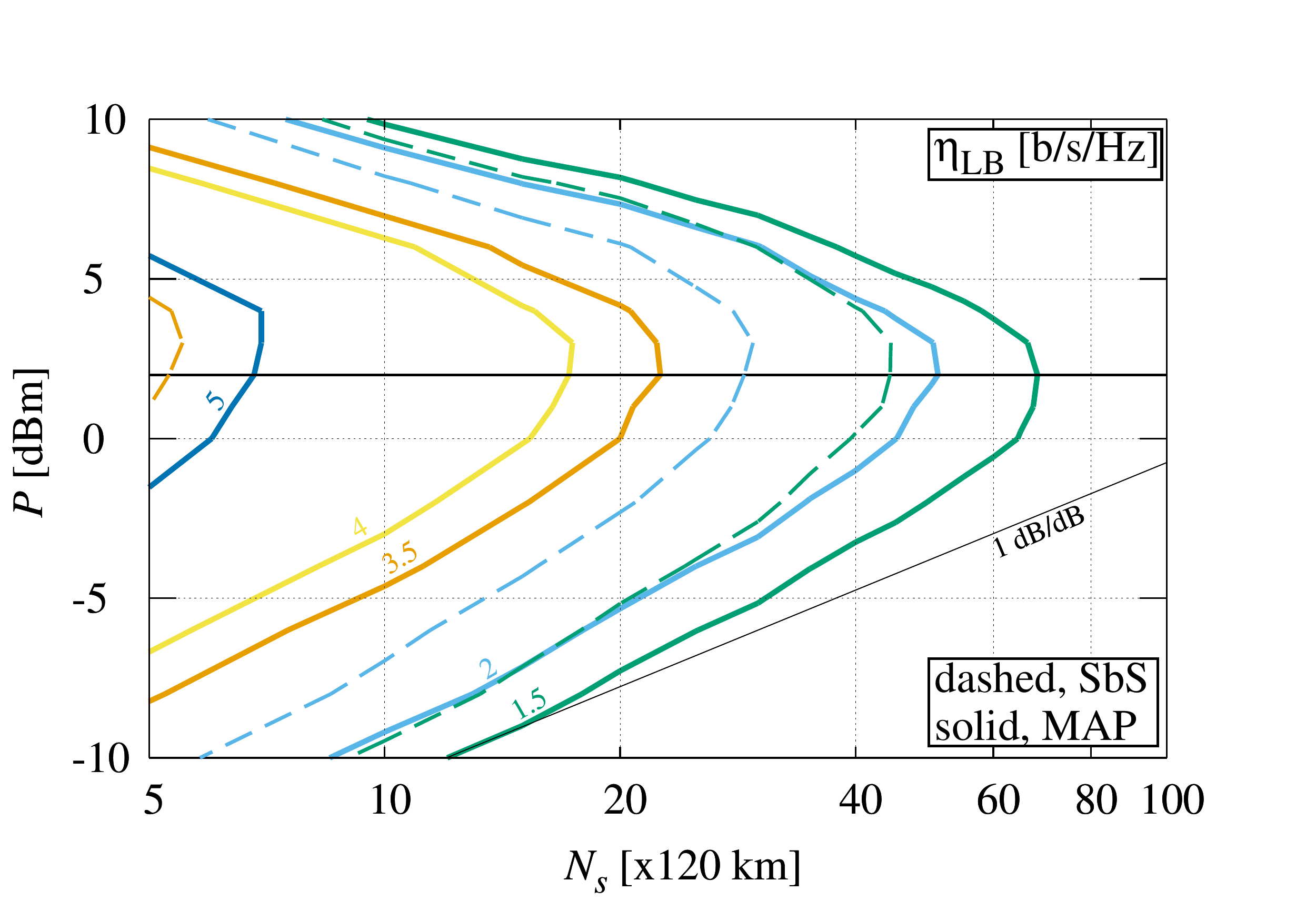}
\par\end{centering}

\protect\caption{\label{fig:grid50}Horizontal cuts of spectral efficiency versus launch
power per channel and number of spans $N_{s}$, in a $N_{s}\times120$
km SMF DU link, with $R$=50 Gbaud, $F$=37.5 GHz, MAP detector with
$L=1$ and conventional symbol-by-symbol detector. }
\end{figure}

We simulated the propagation of $R$=32.5 Gbaud channels, and each
of them was detected by using proper MAP symbol detectors which take
into account a memory of $L=1$ symbols. Since we use a QPSK modulation
per polarization, the detector was split into 4 binary detectors with
$2^{L}=2$ states, each operating on one polarization and one quadrature.
In addition, we also considered the use of a standard symbol-by-symbol
detector. Results are shown in Fig.~\ref{fig:grid32.5}, where horizontal
cuts of SE surfaces are plotted versus $P$ and $N_{s}$. For each
point of the surface we averaged over 6 clusters of about 70000 symbols,
and obtained a confidence interval of at worst 4\%. Furthermore, for
each point the transmitted channels were launched with random initial
polarization, time delays, and offset frequencies. The smoothness
of the resulting surfaces confirms the validity of our averaging.
We also report in this figure the horizontal cut corresponding to
an uncoded bit-error-rate (BER) of 0.0132 (or, equivalently, a Q-factor
equal to 6.95 dB) for the symbol-by-symbol detector, which represents
a conventional pre-FEC BER for a 25\% overhead code (to be more precise,
a concatenated BCH code, see \cite{ITU04}). From the figures, it
can be noticed that the theoretical SE $\eta_{\mathrm{LB}}$ (back-to-back
case) of $\sim4\cdot R/F$ {[}b/s/Hz{]} decreases with increasing
distance, and MAP detector shows a clear maximum reach advantage.
The gain of the MAP detector can be explained with the improved performance
in linear regime, where the limited memory of the channel is properly
exploited by the detector. Such a gain decreases in nonlinear regime
since the ISI introduced by inline filtering is masked by the huge
memory (not accounted for in the auxiliary channel) brought by the
nonlinear channel. Nevertheless, the gain of the MAP detector is still
significant because at the optimal launch power optical noise is twice
as important as nonlinear interference noise \cite{BoRoSe12}. It
is worth noting that the slope of these contour plots in nonlinear
regime are in good agreement with curves shown in \cite{BoRoSe12}
(but it is not equal to 1 dB/dB in linear regime due to inline filtering),
where the Gaussian-noise model for DU optical systems \cite{CaCuBoPoFo12}
was assumed, and signal-to-noise ratio (SNR) cuts are drawn versus
$P$ and $N_{s}$.

\begin{table}
\begin{centering}
\begin{tabular}{|c|c|c|c|c|c|c|c|}
\hline 
\# WSS & 1 & 2 & 5 & 10 & 20 & 50 & 100\tabularnewline
\hline 
Power loss & 8\% & 14\% & 24\% & 32\% & 39\% & 47\% & 53\%\tabularnewline
\hline 
\end{tabular}
\par\end{centering}

\vspace{2mm}

\protect\caption{\label{tab:Power_loss}Signal power loss vs. crossed WSS.}
\end{table}

\begin{figure}
\begin{centering}
\includegraphics[angle=-0,width=0.95\columnwidth]{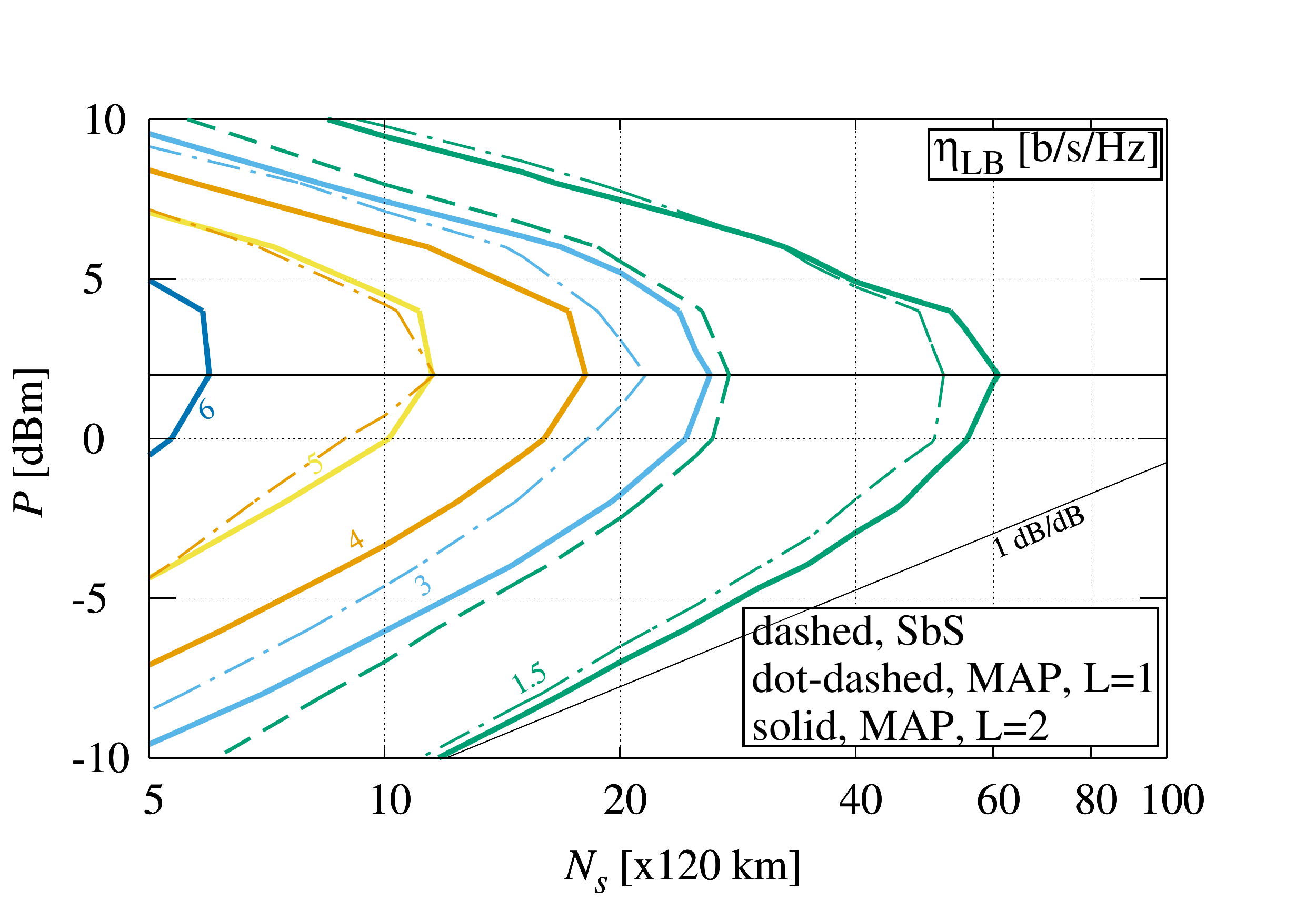}
\par\end{centering}

\protect\caption{\label{fig:grid75}Horizontal cuts of spectral efficiency versus launch
power per channel and number of spans $N_{s}$, in a $N_{s}\times120$
km SMF DU link, with $R$=75 Gbaud, $F$=37.5 GHz, MAP detector with
$L=1$ and $L=2$ and conventional symbol-by-symbol detector. }

\end{figure}

Fig.~\ref{fig:grid50} presents the same horizontal cuts of SE surfaces,
for the case $R$=50 Gbaud. We are here in the realm of time-frequency
packing \cite{BaFeCo09b,CoFo14}. The shape of the contour plots are
very similar to those in the previous figure but, in this case, it
is possible to appreciate the relevant SE improvement (a back-to-back
theoretical value of $\eta_{\mathrm{LB}}\simeq5.35$ {[}b/s/Hz{]}
can be observed at the optimal power), which outlines a clear benefit
with MAP symbol detector especially for short distances. On the contrary,
the threshold detector performs worse than in the $R$=32.5 Gbaud
case already after a few spans, as expected (since it is not able
to cope with the intentional intersymbol interference introduced by
transmit-side narrow filtering). Fig.~\ref{fig:grid75} further shows
results at $R$=75 Gbaud, that is twice the channel spacing. In this
case, since the packing is denser, we also plotted curves for $L=2$
(i.e., four states). A significant SE improvement can be noticed at
shorter distances, where $\eta_{\mathrm{LB}}$ higher than 6 b/s/Hz
is reached. In this scenario, clearly, the symbol-by-symbol dector
performs poorly, reaching at most $\eta_{\mathrm{LB}}\backsimeq1.5$
b/s/Hz.

\begin{figure}
\begin{centering}
\includegraphics[width=0.95\columnwidth]{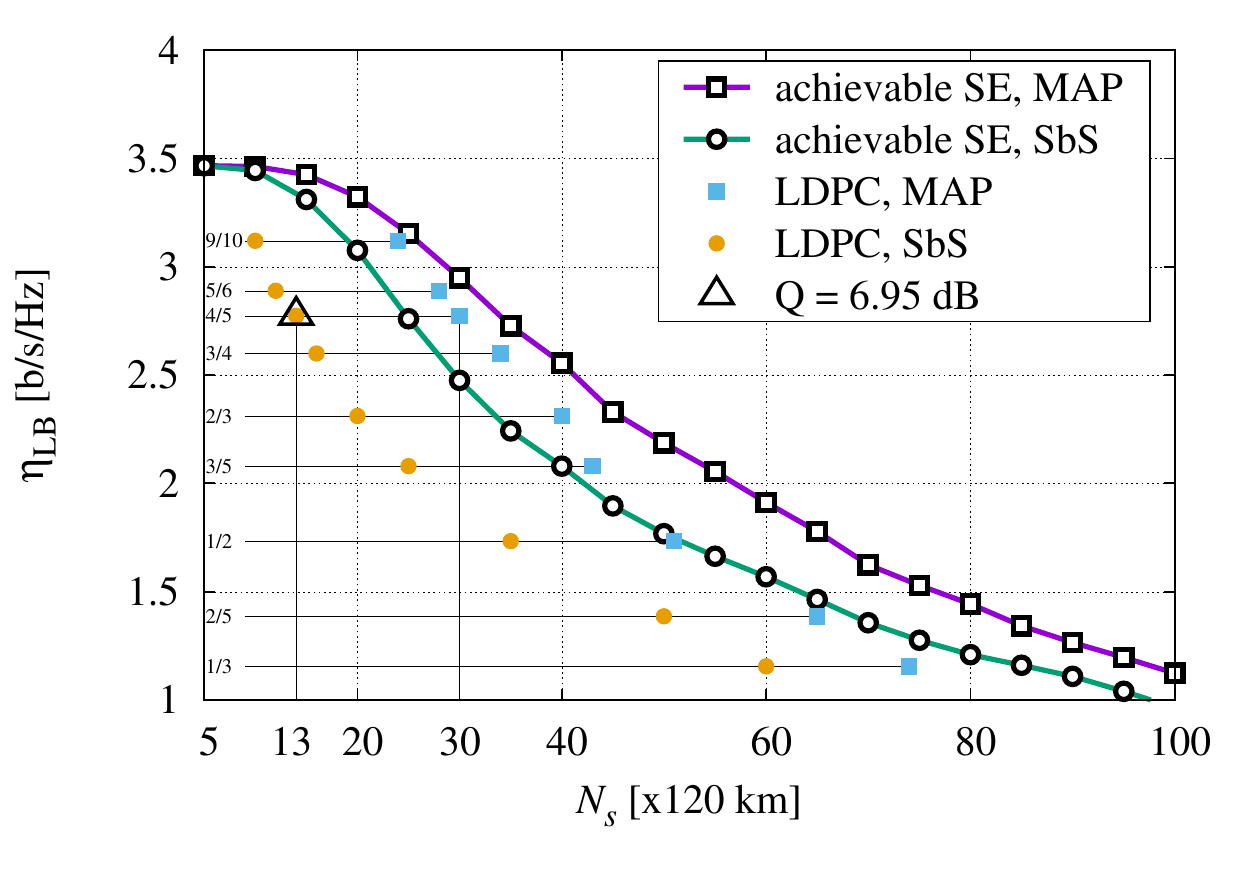}
\par\end{centering}

\protect\caption{\label{fig:P2R32.5}Spectral efficiency versus number of spans, at
2 dBm launch power, $R$=32.5 Gbaud, $F$=37.5 GHz, symbol-by-symbol
(SbS) and MAP detectors, and simulations with rate 1/3, 2/5, 1/2,
3/5, 2/3, 3/4, 4/5, 5/6, 9/10 LDPC codes (with reference BER=$10^{-4}$).
The triangle refers to the pre-FEC Q-factor for a 25\% overhead code.}
\end{figure}

\begin{figure}
\begin{centering}
\includegraphics[width=0.95\columnwidth]{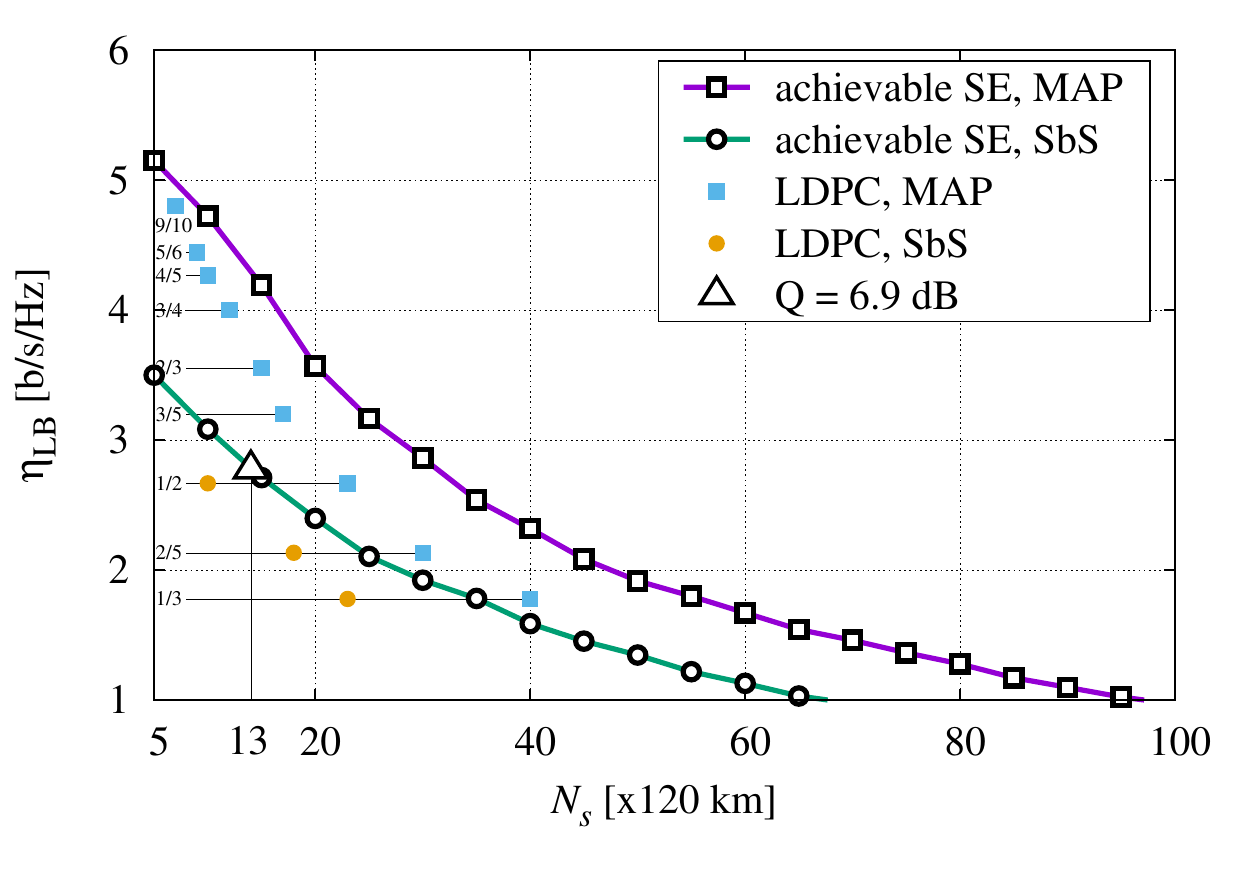}
\par\end{centering}

\protect\caption{\label{fig:P2R50}Spectral efficiency versus number of spans, at 2
dBm launch power, $R$=50 Gbaud, $F$=37.5 GHz, symbol-by-symbol (SbS),
MAP detectors and simulations with rate 1/3, 2/5, 1/2, 3/5, 2/3, 3/4,
4/5, 5/6, 9/10 LDPC codes (with reference BER=$10^{-4}$). The triangle
refers to the pre-FEC Q-factor for a 25\% overhead code.}
\end{figure}

\begin{figure}
\begin{centering}
\includegraphics[width=0.95\columnwidth]{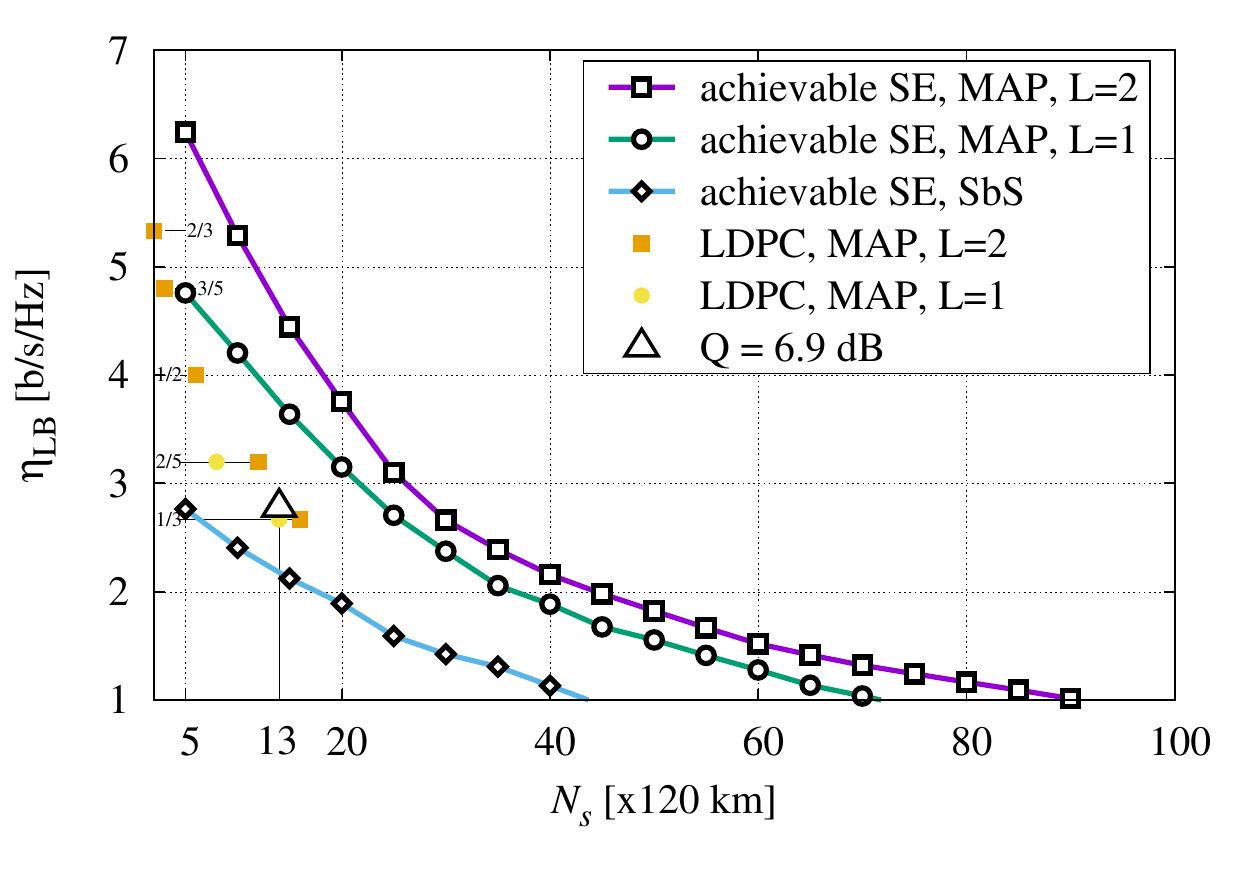}
\par\end{centering}

\protect\caption{\label{fig:P2R75}Spectral efficiency versus number of spans, at 2
dBm launch power, $R$=75 Gbaud, $F$=37.5 GHz, symbol-by-symbol (SbS),
MAP detectors and simulations with rate 1/3, 2/5, 1/2, 3/5, 2/3 LDPC
codes (with reference BER=$10^{-4}$). The triangle refers to the
pre-FEC Q-factor for a 25\% overhead code.}

\end{figure}

Fig.~\ref{fig:P2R32.5}-\ref{fig:P2R75} show instead the vertical
cuts of the SE surfaces at $P$=2 dBm, for $R$=32.5 Gbaud, $R$=50
Gbaud and $R$=75 Gbaud, respectively. For this scenario, we also
simulated coded signaling, by using LDPC codes with rates from 1/3
to 9/10 from \cite{DVB-S2-TR}, and declaring the maximum reach at
distances where an estimated post-FEC BER lower than $10^{-4}$ was
achieved, which in practice implies convergence of the iterative detection/decoding
algorithm.%
\footnote{We consider this value as representative of the error floors of the
simulated LDPC codes, which can be usually found at BER values below
$10^{-7}$. From this BER, additional outer codes with 2-3\% overhead,
tipically BCH, can further reduce the BER down to $10^{-12}$.%
} We fixed a limit of 40 iterations, and averaged over 500,000 received
symbols per step. We found a good agreement between expected results
from achievable lower bounds and simulations, with more affinity for
the MAP detector at small distances, since in this case the auxiliary
channel assumed by the receiver is closer to the true channel (i.e.,
the effect of cascaded ROADMs is still not critical). However, the
gap between achievable SE and simulations becomes larger when the
system is more impaired by ISI and at the same time the actual and
auxiliary channels are more divergent, as can be inferred from Fig.~\ref{fig:P2R75}.
In this case, the system would benefit from a careful re-design of
the employed codes. In any case, simulations with LDPC codes confirm
the reliability of the SE analysis performed through the AIR lower
bounds computation. It is interesting to notice that our simulated
rate-4/5 code with the SbS detector has the same performance as the
6.95-dB reference code indicated by the triangle in Fig.~\ref{fig:P2R32.5},
whereas if the same code is used with the MAP detector the reach can
be more than doubled.

\begin{figure}
\begin{centering}
\includegraphics[width=0.95\columnwidth]{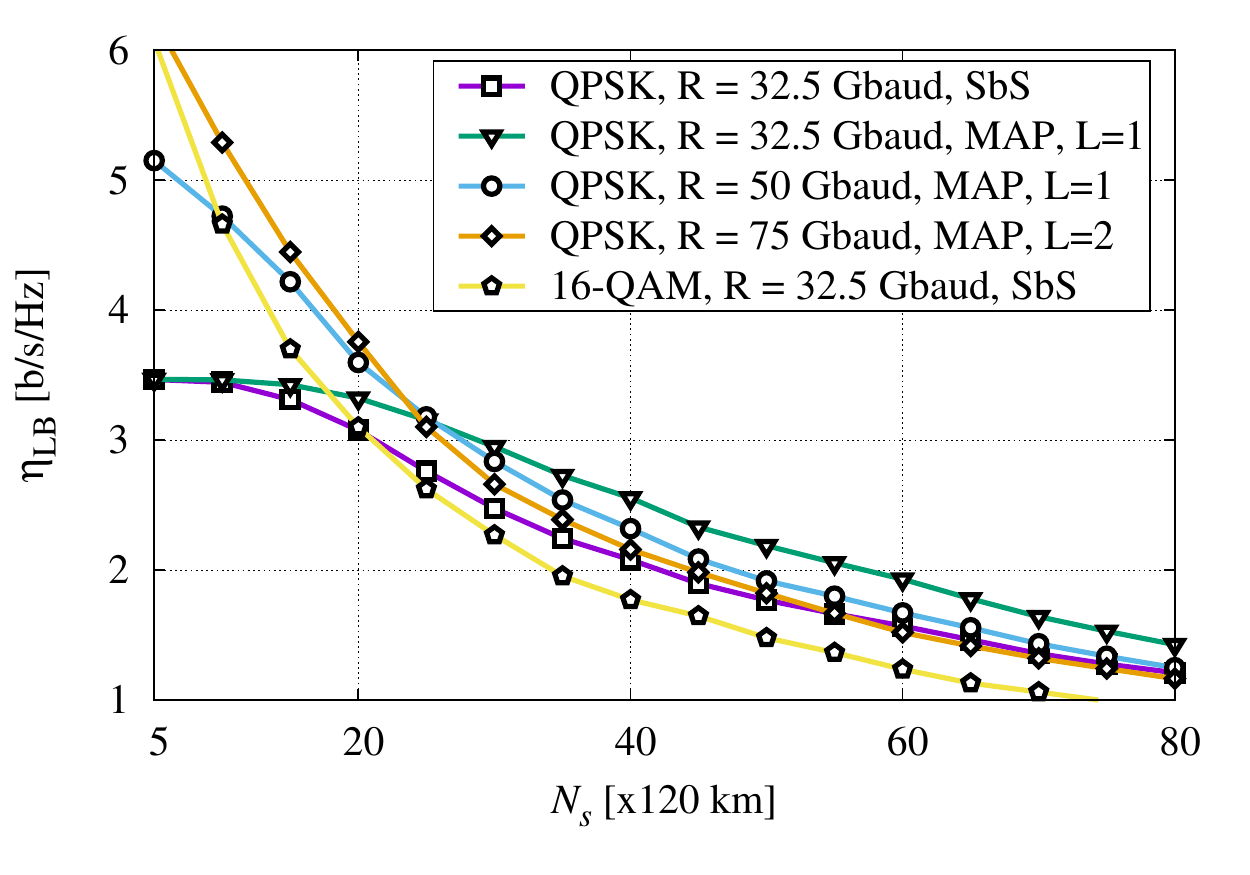}
\par\end{centering}

\protect\caption{\label{fig:P2Sum}Spectral efficiency versus number of spans $N_{s}$,
at 2 dBm launch power, for $R$=32.5 Gbaud, $R$=50 Gbaud and $R$=75
Gbaud, $F$=37.5 GHz, symbol-by-symbol (SbS) and MAP detectors, QPSK
and 16-QAM modulations. }
\end{figure}

Finally, in Fig.~\ref{fig:P2Sum} we report the curves of Fig.~\ref{fig:P2R32.5},
\ref{fig:P2R50} and \ref{fig:P2R75}, in order to provide a direct
and clearer comparison of the results. We also included simulations
of the 16-quadrature amplitude modulation (QAM) format with the symbol-by-symbol
detector at $R=$32.5 Gbaud, in order to provide a more exhaustive
comparison. At equal symbol rate, being the launched signal only slightly
affected by ISI, benefits are remarkable when the effect of inline
WSS filtering starts to significantly impair the signal. On the other
hand, if we consider the 50 and 75 Gbaud signals, an impressive SE
gain up to 80\% is obtained at smaller distances, while there is no
gain after about 25 spans, i.e. 3000 km. 

\begin{figure}
\begin{centering}
\includegraphics[width=0.95\columnwidth]{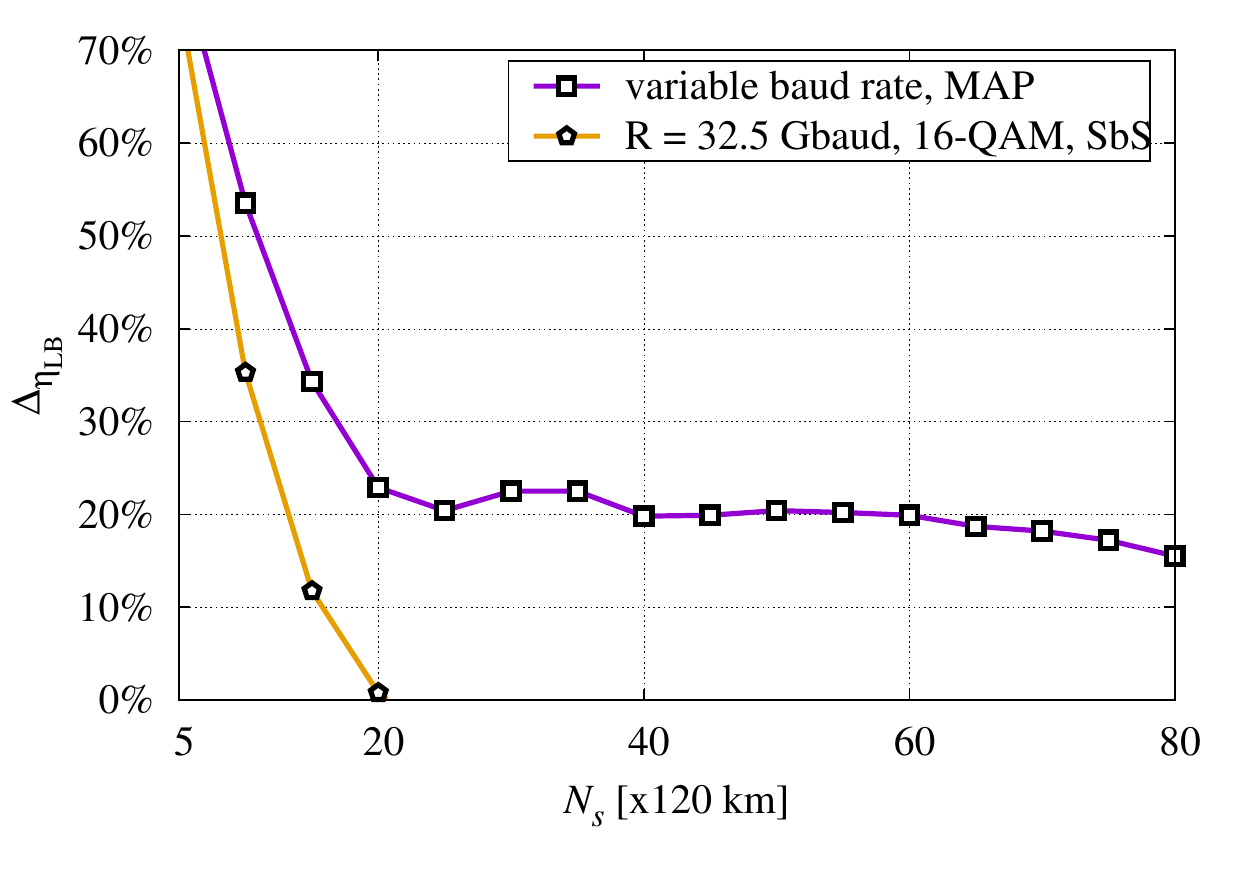}
\par\end{centering}

\protect\caption{\label{fig:P2Delta}Relative spectral efficiency increment with respect
to conventional symbol-by-symbol detector at $R=$32.5 Gbaud, versus
number of spans $N_{s}$. The MAP curve is the envelope of all simulated
symbol rates.}
\end{figure}

By defining the relative SE gain of the MAP receiver with respect
to the SbS receiver as 
\begin{equation}
\Delta_{\eta_{\mathrm{LB}}}=\frac{\eta_{\mathrm{LB}}-\eta_{\mathrm{LB,REF}}}{\eta_{\mathrm{LB,REF}}}\;,\label{eq:Delta}
\end{equation}
where $\eta_{\mathrm{LB,REF}}$ is the achievable SE value of the
QPSK symbol-by-symbol detector with $R$=32.5 Gbaud at each distance,
in Fig.~\ref{fig:P2Delta} we plot the relative gain in SE of MAP
detector versus distance $N_{s}$ in variable symbol rate scenarios
(we simulated 32.5, 37.5, 45, 50, 75 Gbaud channels, in the latest
case with $L=2$), highlighting the benefits of the chosen 2- and
4-state detectors. The MAP curve is given by the envelope of all simulated
symbol rates, so it is possible to infer that the QPSK with MAP detector
performs always better than the 16-QAM, and that choosing a suitable
symbol rate depending on the link length, allows to keep a consistent
SE gain with respect to conventional symbol-by-symbol detector.

\section{Conclusions\label{sec:Conclusions}}

In this paper we investigated the effects of narrow filtering in WDM
transmission over flexible grid optical networks, with particular
emphasis on 37.5 GHz spacing and 32.5 Gbaud rate. We showed that it
is possible to mitigate the detrimental intersymbol interference introduced
by cascading WSS along the link by employing a 2-state maximum-a-posteriori
symbol detector for each input bit stream. The mitigation is quantified
in terms of achievable spectral efficiency versus propagation distance
and launch power. Moreover, we proved that it is possible to exploit
LDPC codes of different rates in order to achieve a desired spectral
efficiency, thus obtaining a remarkable system reach improvement,
or, conversely, higher spectral efficiency at equal reach, with respect
to the conventional symbol-by-symbol detector. Finally, we proved
that by transmitting 50 Gbaud channels on the same DU SMF link a great
SE gain can be achieved by working beyond the Nyquist-WDM limit through
time-frequency packing, which enables a remarkable spectral efficiency
gain for distances up to 3000 km.

\section*{Acknowledgment}

This work was supported in part by CNIT and by the Italian Ministero
dell'Istruzione, dell'Università e della Ricerca (MIUR) under the
FIRB project Coherent Terabit Optical Networks (COTONE).

\bibliographystyle{IEEEtran}
\bibliography{Refs,Refs_TF}

\begin{thebibliography}{10}
\providecommand{\url}[1]{#1}
\csname url@samestyle\endcsname
\providecommand{\newblock}{\relax}
\providecommand{\bibinfo}[2]{#2}
\providecommand{\BIBentrySTDinterwordspacing}{\spaceskip=0pt\relax}
\providecommand{\BIBentryALTinterwordstretchfactor}{4}
\providecommand{\BIBentryALTinterwordspacing}{\spaceskip=\fontdimen2\font plus
\BIBentryALTinterwordstretchfactor\fontdimen3\font minus
  \fontdimen4\font\relax}
\providecommand{\BIBforeignlanguage}[2]{{%
\expandafter\ifx\csname l@#1\endcsname\relax
\typeout{** WARNING: IEEEtran.bst: No hyphenation pattern has been}%
\typeout{** loaded for the language `#1'. Using the pattern for}%
\typeout{** the default language instead.}%
\else
\language=\csname l@#1\endcsname
\fi
#2}}
\providecommand{\BIBdecl}{\relax}
\BIBdecl

\bibitem{BoCuCaPoFo11}
G.~Bosco, V.~Curri, A.~Carena, P.~Poggiolini, and F.~Forghieri, ``On the
  performance of {N}yquist-{WDM} terabit superchannels based on {PM}-{QPSK},
  {PM}-{8PSK} or {PM}-16{QAM} subcarriers,'' \emph{J.~Lightw. Tech.}, vol.~29,
  no.~1, pp. 53--61, 2011.

\bibitem{jansen:OFT09}
S.~L. Jansen, B.~Spinnler, I.~Morita, S.~Randel, and H.~Tanaka, ``{100GbE: QPSK
  versus OFDM},'' \emph{Optical Fiber Technology}, vol.~15, no. 5-6, pp.
  407--413, Oct-Dec 2009.

\bibitem{Chraplyvy10}
R.~Tkach and A.~Chraplyvy, ``Past/present system advances, with an eye towards
  the future,'' in \emph{Proc. European Conf. on Optical Commun. ({ECOC}'10)},
  Sep. 2010.

\bibitem{ITU12}
{ITU-T}{ G.694.1}, ``Spectral grids for {WDM} applications: {DWDM} frequency
  grid,'' Feb. 2012.

\bibitem{Finisar12}
Finisar, ``Balancing performance, flexibility, and scalability in optical
  networks,'' White Paper, Feb. 2012.

\bibitem{MoreaOFC14}
A.~Morea, J.~Renaudier, A.~Ghazisaeidi, O.~Bertrand-Pardo, and T.~Zami,
  ``Impact of reducing channel spacing from 50 {GHz} to 37.5 {GHz} in fully
  transparent meshed networks,'' in \emph{Proc.~{O}ptical {F}iber {C}ommun.
  {C}onf.}, San Francisco, CA, 2014, paper Th1E.4.

\bibitem{BaCoJeRa74}
L.~R. Bahl, J.~Cocke, F.~Jelinek, and J.~Raviv, ``Optimal decoding of linear
  codes for minimizing symbol error rate,'' \emph{IEEE Trans.~Inform.~Theory},
  vol.~20, pp. 284--287, Mar. 1974.

\bibitem{CoFo14}
G.~Colavolpe and T.~Foggi, ``Time-frequency packing for high capacity coherent
  optical links,'' \emph{IEEE Trans. Commun.}, vol.~62, no.~8, pp. 2986--2995,
  Aug. 2014.

\bibitem{DVB-S2-TR}
\BIBentryALTinterwordspacing
{ETSI EN 301 307 Digital Video Broadcasting (DVB); V1.1.2 (2006-06)}, ``{\rm
  Second generation framing structure, channel coding and modulation systems
  for Broadcasting, Interactive Services, News Gathering and other Broadband
  satellite applications}.'' [Online]. Available: \url{http://www.etsi.org}
\BIBentrySTDinterwordspacing

\bibitem{CoFoMoPi11}
G.~Colavolpe, T.~Foggi, A.~Modenini, and A.~Piemontese, ``Faster-than-{N}yquist
  and beyond: how to improve spectral efficiency by accepting interference,''
  \emph{Opt. Express}, vol.~19, no.~27, pp. 26\,600--26\,609, Dec 2011.

\bibitem{Men03}
O.~V. Sinkin, R.~Holzl\"ohner, J.~Zweck, and C.~R. Menyuk, ``Optimization of
  the split-step fourier method in modeling optical-fiber communications
  systems,'' \emph{J.~Lightw. Technol.}, vol.~21, no.~1, pp. 61--68, Jan. 2003.

\bibitem{CoFoFoPr08}
G.~Colavolpe, T.~Foggi, E.~Forestieri, and G.~Prati, ``Multilevel optical
  systems with {MLSD} receivers insensitive to {GVD} and {PMD},''
  \emph{J.~Lightw. Tech.}, vol.~26, pp. 1263--1273, May 2008.

\bibitem{RuPr12}
F.~Rusek and A.~Prlja, ``Optimal channel shortening for {MIMO} and {ISI}
  channels,'' \emph{IEEE Trans. Wireless Commun.}, vol.~11, no.~2, pp.
  810--818, Feb. 2012.

\bibitem{ArLoVoKaZe06}
D.~M. Arnold, H.-A. Loeliger, P.~O. Vontobel, A.~Kav\v{c}i\'c, and W.~Zeng,
  ``Simulation-based computation of information rates for channels with
  memory,'' \emph{IEEE Trans.~Inform.~Theory}, vol.~52, no.~8, pp. 3498--3508,
  Aug. 2006.

\bibitem{EssTka12}
R.~Essiambre and R.~W. Tkach, ``Capacity trends and limits of optical
  communication networks,'' \emph{Proc. IEEE}, vol. 100, no.~5, pp. 1035--1055,
  May 2012.

\bibitem{Essjlt10}
R.~Essiambre, G.~Kramer, P.~J. Winzer, G.~J. Foschini, and B.~Goebel,
  ``Capacity limits of optical fiber networks,'' \emph{J.~Lightw. Technol.},
  vol.~28, no.~4, pp. 662--701, Feb. 2010.

\bibitem{MeKaLaSh94}
N.~Merhav, G.~Kaplan, A.~Lapidoth, and S.~Shamai, ``On information rates for
  mismatched decoders,'' \emph{IEEE Trans.~Inform.~Theory}, vol.~40, no.~6, pp.
  1953--1967, Nov. 1994.

\bibitem{Serena12}
P.~Serena, A.~Bononi, and G.~Colavolpe, ``On the nonlinear capacity with memory
  of {PS}-{QPSK} and {PDM}-{QPSK} in {WDM} non-dispersion managed links,'' in
  \emph{Proc. European Conf. on Optical Commun. ({ECOC}'12)}, Sep. 2012, paper
  We.2.C.1.

\bibitem{CoBa05b}
G.~Colavolpe and A.~Barbieri, ``On {MAP} symbol detection for {ISI} channels
  using the {U}ngerboeck observation model,'' \emph{IEEE Commun. Letters},
  vol.~9, no.~8, pp. 720--722, Aug. 2005.

\bibitem{ITU04}
{ITU-T}{ G.675.1}, ``Forward error correction for high bit-rate {DWDM}
  submarine systems,'' Feb. 2004.

\bibitem{BoRoSe12}
A.~Bononi, N.~Rossi, and P.~Serena, ``On the nonlinear threshold versus
  distance in long-haul higly-dispersive coherent systems,''
  \emph{Opt.~Express}, vol.~20, no.~26, pp. 204--216, Dec. 2012.

\bibitem{CaCuBoPoFo12}
A.~Carena, V.~Curri, G.~Bosco, P.~Poggiolini, and F.~Forghieri, ``Modeling of
  the impact of nonlinear propagation effects in uncompensated optical coherent
  transmission links,'' \emph{J.~Lightw. Technol.}, vol.~30, no.~10, pp.
  1524--1539, May 2012.

\bibitem{BaFeCo09b}
A.~Barbieri, D.~Fertonani, and G.~Colavolpe, ``Time-frequency packing for
  linear modulations: spectral efficiency and practical detection schemes,''
  \emph{IEEE Trans. Commun.}, vol.~57, pp. 2951--2959, Oct. 2009.

\end{thebibliography}

\end{document}